# Towards Designing PLC Networks for Ubiquitous Connectivity in Enterprises

Kamran Ali[†]  Ioannis Pefkianakis[‡]  Alex X. Liu[†]  Kyu-Han Kim[‡]
[†]Dept. of Computer Science and Engineering, Michigan State University, USA
[‡]Hewlett Packard Enterprise Laboratories (HPE Labs), Palo Alto, California, USA
E-mail: [†]{alikamr3, alexliu}@cse.msu.edu, [‡]{ioannis.pefkianakis, kyu-han.kim}@hpe.com

*Abstract*—Powerline communication (PLC) provides inexpensive, secure and high speed network connectivity, by leveraging the existing power distribution networks inside the buildings. While PLC technology has the potential to improve connectivity and is considered a key enabler for sensing, control, and automation applications in enterprises, it has been mainly deployed for improving connectivity in homes. Deploying PLCs in enterprises is more challenging since the power distribution network is more complex as compared to homes. Moreover, existing PLC technologies such as HomePlug AV have not been designed for and evaluated in enterprise deployments. In this paper, we first present a comprehensive measurement study of PLC performance in enterprise settings, by analyzing PLC channel characteristics across space, time, and spectral dimensions, using commodity HomePlug AV PLC devices. Our results uncover the impact of distribution lines, circuit breakers, AC phases and electrical interference on PLC performance. Based on our findings, we show that careful planning of PLC network topology, routing and spectrum sharing can significantly boost performance of enterprise PLC networks. Our experimental results show that multi-hop routing can increase throughput performance by 5x in scenarios where direct PLC links perform poorly. Moreover, our trace driven simulations for multiple deployments, show that our proposed fine-grained spectrum sharing design can boost the aggregated and per-link PLC throughput by more than 20% and 100% respectively, in enterprise PLC networks.

## I. INTRODUCTION

**Motivation & Background:** Power line communication (PLC) technology has gained popularity as a connectivity solution in homes and various smart grid related applications. The HomePlug Powerline Alliance is leading the standardization efforts, with over 100 million HomePlug devices in the market and annual growth rate of over 30% [1]. PLC devices nowadays can support high-bandwidth applications such as HD video streaming and VoIP, while boasting rates greater than 1 Gbps [2], [3]. Although PLCs have been widely adopted in home settings for extending LAN/WLAN network coverage and interconnecting home computers, peripherals, entertainment devices and sensors, the deployment of PLCs in large buildings such as enterprises remains largely unexplored. One reason for the limited industrial-scale PLC deployment is the concern that PLC performance will not scale as more PLC nodes are added in the network [4].

PLC technology has the potential to provide high-speed ubiquitous connectivity, and facilitate new applications in large-building settings (such as enterprises) at low cost, without any need for dedicated network cabling. First, PLC can augment existing Wi-Fi enterprise networks [5], to provide connectivity to wireless blind spots and to accommodate traffic from overloaded Access Points. PLC is further considered as a key enabler for sensing, control, and automation in large-scale smart grid systems, which comprise of hundreds of sensors spread over wide areas. Specifically, PLC can connect energy meters and SCADA (Supervisory Control and Data Acquisition) sensors to the smart grid control center. In the context of IoT, PLC can provide a simple and cost effective back haul connectivity to sensors deployed at different parts of an enterprise [4].

**Challenges of Enterprise PLC Networks:** The deployment of PLC networks in large buildings such as enterprises is more challenging than home network settings for three reasons. First, the power distribution networks in enterprises are more complex and comprise higher dynamics as compared to those in homes. In enterprise buildings there are typically multiple power distribution lines coming in from main switch boards (MSBs). Each distribution line consists of 3 AC phases, which are further distributed into the building through multiple breaker circuits. Such network components can significantly affect PLC performance. Second, PLC channel dynamics, which can be attributed to a multitude of electrical devices connected to power lines, can be significant and highly location dependent in enterprises. Third, the deployment of PLCs in large buildings requires the deployment of multiple PLC devices to provide ubiquitous coverage, which can lead to high inter-link interferences. Routing and spectrum allocation for improved connectivity in such scenarios remain unexplored.

**Measurement Approach:** In this paper, we first present an extensive and in-depth measurement study of PLC network performance in enterprise settings, using commodity HomePlug AV (HPAV) hardware. We leverage open source PLC software tools, which allow us to extract fine PHY-layer and MAC-layer feedback such as tonemaps (i.e. per-subcarrier modulation) from HPAV devices. Our study departs from recent measurement efforts [4]–[7] in two ways. First, we characterize the impact of the power distribution network components (distribution lines, AC phases, circuit breakers) of large buildings, on PLC performance. We further isolate the impact of interfering electrical appliances on PLCs. Second, our analysis captures the *a) spatial*, *b) temporal*, and *c) spectral* (per OFDM subcarrier level) dynamics of PLC networks in enterprise settings. Finally, based on the results from our measurement study, we explore multi-hop routing in PLCs and propose a novel spectrum sharing scheme for PLCs.

**Findings & Solutions:** Our study uncovers several important findings related to the behavior of PLC networks in enterprise settings. We first observe that the communication among PLC nodes located at different power distribution lines is often impossible. Consequently, large buildings consist of multiple disconnected PLC networks. For PLC nodes located under the same distribution line, AC phases do not have a significant effect on PLC performance, while circuit breakers can drop PLC throughput by 20%-30%. Moreover, interfering electrical appliances are a key factor of performance degradation in enterprise PLC networks. Such power distribution network characteristics and interfering appliances result into temporal dynamics and asymmetries in PLC channels. Specifically, we observe performance variations across all the OFDM subcarriers of the communication spectrum, even for PLC nodes in the same neighborhood. Although our measurements were conducted with HPAV devices, our results can be generalized for other broadband PLC technologies.

Our experimental findings raise important, still open questions related to the deployment of PLC networks in large building settings. How to design PLC network topology? How to share the available spectrum among PLC nodes more efficiently? Can PLC routing improve performance? To this end, we first propose a PLC network (*PLC-Net*) topology, which can provide ubiquitous PLC connectivity in enterprise settings. We then evaluate routing in PLC-Net, and we show using testbed experiments that multi-hop routing (such as OLSR routing adopted in 802.11 wireless mesh networks [8]) can boost connectivity, and can provide more than 5× throughput gains in certain scenarios. Finally, as HPAV networks currently do not support any spectrum allocation strategies, there can be scenarios where a subset of subcarriers are highly underutilized. Our finding that per-subcarrier performance varies among neighboring links in PLCs makes case for a dynamic spectrum sharing solution, where the low-modulated subcarriers of a certain PLC link can be utilized by other links, to improve the aggregated network performance. We propose and evaluate such a spectrum sharing approach and show the performance gains in terms of throughput and fairness.

## II. Related work

There are several areas of research related to our work. We briefly comment on them and position our contributions.

The performance of PLC in larger size buildings has been studied previously through analysis and trace-driven simulations. Authors in [9] investigate the impact of domestic breakers on low voltage power line communications. They find that, the impact of some breakers is more considerable in broadband ([1-100] MHz) than narrowband ([9-500] kHz). Authors in [10] show that the exact signal power attenuation is proportional to the number of branch circuits in the distribution boards. They further propose a single wire PLC solution to shorten the communication distance and to reduce the attenuation, which requires the installation of new wiring. In [11], authors further show that the attenuation of branch lines connected to a distribution board is 20 dB, at the MHz bands. Different from the above efforts, we conduct measurements using commodity HPAV devices and study all aspects (i.e,. spatial, temporal and spectral) of the performance of PLC networks in enterprise settings.

Recent studies evaluate PLC performance using HPAV commodity testbeds. In [5], [7] authors compare HPAV with WiFi performance. They study temporal and spatial variations of the throughput of PLC links and make a case for hybrid PLC-WiFi networks [5]. The measurement study in [6] shows the multi-flow performance of PLC networks. It then presents BOLT, which seeks to manage traffic flows in PLC networks. Different from our work, the above studies mainly focus on temporal variations and the impact of connected electrical devices on PLC performance, without considering the power distribution network components (breakers, phases, distribution lines), which highly determine PLC performance in an enterprise setting. However, the analysis in previous studies [12], [13] shows that throughputs in PLC networks decrease with number of contending stations. This is the main motivation behind our novel spectrum sharing strategy for PLCs, which has the potential to enhance the throughput as well as fairness in enterprise level PLC deployments, while incurring minimum changes to current HPAV protocols.

In [14], [15], authors evaluate the performance of existing geo-routing protocols in PLC networks using simulations. In [16], authors discuss an extension of LOAD routing protocol to make it suitable for PLC networks and perform a limited evaluation of their proposed protocol. However, all these efforts are either related to PLCs over electric grids [14], [15] or do not take into account practical issues such as temporal variations due to electrical appliances and the impact of power distribution network components. To our knowledge, our study is the first to evaluate routing in PLC networks using real testbed experiments. Moreover, previous work does not take into account spectral properties of PLC channels while studying medium access and routing in PLCs. To our knowledge, there is no prior research study on bandwidth sharing in PLCs.

In summary, different from the aforementioned works, we extensively study the behavior of PLCs in spatial, temporal and spectral dimensions, using commodity PLC devices. We further provide guidelines to design enterprise PLC networks by discussing PLC network topology design and possible advantages of multi-hop routing. Finally, we propose a novel bandwidth sharing strategy for PLCs which can improve fairness and aggregated throughput of PLC-Nets.

## III. PLC Dataset

In this section, we first describe the main characteristics of PLC networks related to our study, focusing on the widely deployed HomePlug AV (HPAV) IEEE 1901 [17] standard. Afterwards, we explain our data collection approach.

### A. PLC Background

*1) PLC Channel Characteristics:* A large body of work on PLC has focused on theoretical modeling of PLC channel characteristics [18]–[24]. Multipath is a key characteristic of PLC channels, which is attributed to unmatched loads or branch circuits connected to different sockets on the powerline.

In a typical power distribution network of a large building, there are multiple branch circuits with different impedances, and therefore, PLC signals are reflected from multiple reflection points leading to multipath effects. On top of multipath attenuations, several different types of noise in PLC channels have been identified [23], [24]. Harmonics of AC mains and other low power noise sources in the power lines lead to colored background noise, which decreases with frequency. This noise is usually negligible in PLCs operating in MHz bands. Radio broadcasts in AM bands create narrow band noise consisting of sinusoidal signals with modulated amplitudes. Periodic impulsive noise is created due to rectifiers, switching power supplies and AC/DC converters, which can be either synchronous or asynchronous with AC line cycle. Aperiodic impulsive noise also exists in PLC channels mainly due to switching transients in electric power supplies, AC/DC converters, etc.

*2) HomePlug AV standard:* The most widely adopted family of PLC standards are HomePlug AV, AV2 and Green PHY standards [25]. HomePlug AV2, which is the latest of these standards, can support up to 1 Gbps PHY rates. Our study focuses on the HomePlug AV standard, which has been widely used in home networks to improve coverage, and can support maximum PHY rates of up to 200 Mbps [2], [3].

**PHY-layer:** One of the main difference between PLC devices and WiFi devices is that PLCs use the whole frequency band for communication and WiFi-like channelization is not possible. HPAV uses 1.8-30 MHz frequency band and employs OFDM with 917 subcarriers, where each subcarrier can use any modulation scheme from BPSK to 1024-QAM depending on the channel conditions [25]. In order to update the modulation schemes for each subcarrier, two communicating HPAV PLC devices continuously exchange and maintain *tonemaps* between them. Tonemaps refer to the information about the modulation scheme used per subcarrier, i.e. the number of bits modulated per subcarrier. The tonemaps exchanged are estimated for multiple different sub-intervals of the AC mains cycle. Tonemaps are exchanged between PLC devices through a *sounding* process, where the transmitter sends sounding frames to the receiver using QPSK for all subcarriers, the destination estimates the channel quality and sends back the tonemaps corresponding to different sub-intervals of AC mains cycle back to the transmitter. The destination can communicate up to 7 tonemaps, i.e. 6 tonemaps for the different sub-intervals of the AC line cycle called slots and one default tonemap [25]. Tonemaps are continuously updated by default after 30 seconds or when the error rate exceeds a threshold [25]. *Tonemaps provide us with indirect information about Channel Frequency Response (CFR) between communicating PLC devices.*

**MAC-layer:** Both TDMA and CSMA/CA are supported by HPAV [25]. The CSMA protocol of HPAV devices is different from 802.11 CSMA/CA used by WiFi devices. HPAV devices increase their contention windows not only after a collision, but also after sensing the medium to be busy [12]. The Request to Send (RTS) and Clear to Send (CTS) delimiters can be enabled in HPAV stations during CSMA slots, to handle hidden nodes. HPAV frames are 512 byte aggregated physical blocks (PBs) of data. Reception of each PB of a frame is separately acknowledged, so that the transmitter retransmits only the PBs which are corrupted, either due to collisions or channel variations.

*B. Data Collection*

**Experimental setup:** Our study is based on measurements with commodity HomePlug AV hardware. We use Meconet HomePlug AV mini-PCI adapters with Intellon INT6300 chipsets, which can support 200 Mbps PHY rates. We connect the PLC adapters to ALIX 2D2 boards, which run OpenWrt operating system. We use open source PLC software tool named *open-plc-utils*, which is developed by Qualcomm, to extract PHY and MAC-layer feedback (such as tonemaps), directly from the Meconet HPAV adapters.

**Experimental methodology:** For our experiments we place our PLC nodes in various locations in the floorplan of Figure 1, of an enterprise building. We elaborate on the power distribution network of our building in the following sections. We generate saturated *iperf* UDP traffic among the PLC nodes (unless it is explicitly stated). The results reported in this paper are averaged over multiple runs.

**Metrics:** We analyze the performance of PLC networks by first collecting iperf throughput and jitter statistics. We further elaborate on the per-subcarrier PLC network performance by analyzing the tonemaps extracted by the open-plc-utils software tool running on PLC nodes. For a given PLC communication link and for the $k^{th}$ sub-interval of AC line cycle, the effective PHY rate can be estimated from tonemaps as follows [17]:

$$R_{phy}^{\{k\}} = \frac{[\sum_{j=1}^{N} T[j]^{\{k\}}] \cdot C^{\{k\}} \cdot (1 - B_{err}^{\{k\}})}{T_s} \quad (1)$$

where $j$ is subcarrier number and $N$ is total number of subcarriers. $T_j$ is the modulation rate (i.e., bits per subcarrier) of the $j^{th}$ subcarrier. $C$ is Forward Error Correction (FEC) code rate. HomePlug AV supports FEC code rates of 1/2 and 16/21. Finally, $B_{err}$ is the bit error rate and $T_s$ is the symbol interval of OFDM communication. $T_s$ is approximately ~46$\mu s$ for HomePlug AV including all overheads [25]. The expected throughput, averaged over all the sub-intervals of the AC line cycle, can be approximated as:

$$Thr \approx (1 - F_o) \cdot \frac{\sum_{k=1}^{N_{AC}} R_{phy}^{\{k\}}}{N_{AC}} \quad (2)$$

where $F_o$ accounts for HPAV protocol overheads and $N_{AC}$ is the number of sub-intervals of AC line cycle. $N_{AC}$ is 5 or 6 for USA frequencies and $F_o$ is typically $\sim 0.4$ based on our experiments. In rest of the paper, we report iperf throughput measurements and tonemaps. In all our experiments, we observed that the FEC code rate was always 16/21 for the communication among our HPAV devices. Therefore, based on equations 1, 2, the tonemaps of a PLC link reflect its throughput performance.

IV. EFFECTS OF POWER DISTRIBUTION NETWORK

In this section, we present our analysis of the measurements we conducted to study the impact of different components of a power distribution network (e.g. phases, breakers and distribution/trunk lines) on PLC performance in an enterprise building setting. We measure the throughput and jitter performance of more than 40 links (PLC transmitter-receiver pairs) for the analysis of PLC links in aforementioned scenarios.

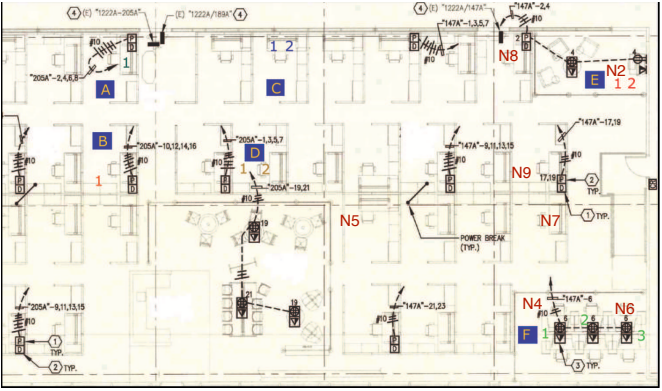

Fig. 1: Building power distribution plan.

*A. Overview of Power Distribution Network*

An overview of the power distribution network floorplan of the enterprise building in which we conducted our experiments is shown in Figure 1. The main switchboard of the enterprise steps down the voltage from thousands to hundreds of Volts and the down converted electric power is then distributed towards different floors of different buildings in the enterprise through multiple different distribution lines or *trunk lines* (represented with hexagonal boxes with #4 written on them). The power from the trunk lines coming into the floor of a building is then further distributed into different parts of the floor through a distribution board containing a set of circuit breakers which divide the electrical power feed into subsidiary circuits. Each trunk line consists of 3 cables corresponding to 3 different phases and each distribution board contains multiple breakers per phase. This is a typical power distribution plan of most enterprise and other campus buildings in USA[1].

The letters and numbers in the floorpan of Figure 1 represent some of the different locations we place our PLC nodes. Next, we elaborate on our experiments.

*B. Performance for Same Phase & Breaker*

We first evaluate the PLC network performance for nearby nodes, while they operate on the same breaker, phase and distribution line. We expect such a setting to give us the upper bound of PLC performance, since there are no attenuations related to the power distribution network in this scenario. We measure the throughput and jitter performance of 18 links in this scenario. The blue line in figure 2 shows the CDF plot of the average throughputs measured over all 18 links.

The maximum throughput observed is 88.6 Mbps, while the peak instantaneous throughput achieved by our nodes, never exceeds 90 Mbps. Moreover, as shown by the CDF in figure 2, throughputs of more than 70 Mbps were observed across the tested links approximately 75% of the time. Jitter is overall low, with the median being 0.2 ms and a maximum of 2.5 ms. We observe these high throughputs when there are no interfering devices in proximity of the PLC nodes corresponding to a PLC link. This can be observed through tonemaps, which shows most of the subcarrier to be fully modulated. For example, Figures 3(a) and 3(b) show the tonemaps for two different links (N1-N2 and N2-N1) for 5 different sub-intervals of AC line cycle. Both nodes are connected to phase B (#3 on 205-A trunk line in figure 1).

---

[1]http://electrical-engineering-portal.com/north-american-versus-european-distribution-systems

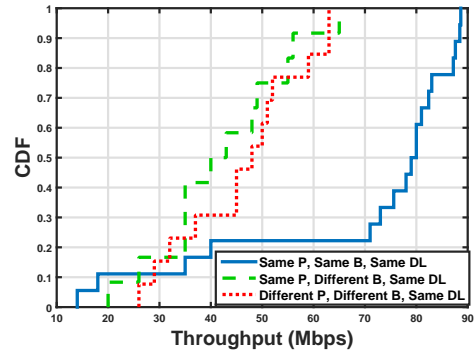

Fig. 2: CDF of throughputs observed in different cases.

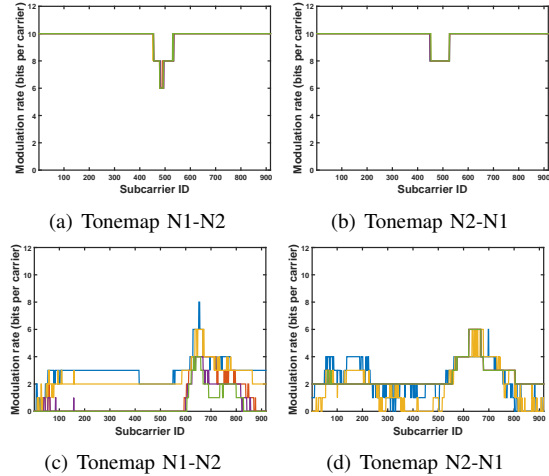

Fig. 3: Tonemaps for same phase, breaker, trunk line (a)-(b) without interference, (c)-(d) with interference.

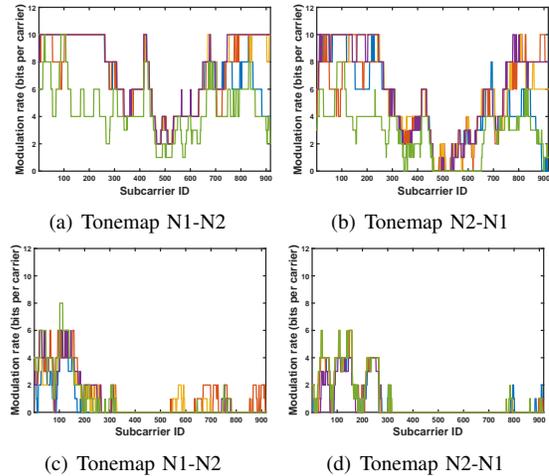

Fig. 4: Tonemaps for (a)-(b) different breakers and (c)-(d) different distribution lines.

The low performance observed during our experiments is attributed to interfering devices, which generate in-band interference in PLC nodes and introduce high multipath attenuations, which depend on the electric load (switching circuitry and impedance) of the devices. Figures 3(c) and 3(d) show the tonemaps for the worst performing PLC link, which achieved 14 Mbps throughput and 2.5 ms jitter. In this scenario, nodes placed in locations F-1, F-3 were connected to phase C (#6 on 147-A distribution line shown in 1) and there were several devices like laptops, smartphones, LED monitors, connected to nearby sockets (all devices were connected to the same phase and breaker). We can observe that most of the subcarriers have very low or even zero bits modulation. We further elaborate on the impact of connected electric devices on PLC performance, in Section V-B while discussing dynamics of PLC channel.

**Conclusion 1:** *The performance of a PLC links operating on same breaker and same distribution line is mainly affected by the location of PLC nodes with respect to the interfering electrical appliances. Highly attenuating device impedances or severe device interferences can lead to significant performance degradation (we observe ~6.5 fold decrease in throughput).*

### C. Impact of Breakers

Next, we analyze the impact of breakers on PLC performance. We evaluate the throughput and jitter performance of multiple PLC links, where transmitter-receiver pairs are connected under different breakers, but on the same phase and distribution line. We observe maximum throughput of 63 Mbps, which is 25.6 Mbps lower (29% decrease) than the previous case where nodes were connected under the same breaker. The median throughput observed was 51 Mbps, with minimum being 26 Mbps, which is higher than the minimum of same breaker case as we did not encounter a high interference case in this scenario. The green line in Figure 2 shows the CDF plot of the throughputs observed.

Our results show that breakers can affect certain subcarriers more than others. For example, Figure 4(a) and 4(b) show the tonemaps between two links for the 5 different sub-intervals of the AC line cycle. Node 1 located at B is connected to phase B (#16 on 205-A distribution line in Figure 1) and node 2 located at D is also connected to phase B (#3 on 205-A distribution line in Figure 1). In this experiment, we observed average throughputs of 50 Mbps and 52 Mbps in both directions respectively. We also observe modulations varying between zero to 10 bits for different subcarriers in different AC line cycle slots. These variations are attributed to both breakers and other electric appliances connected in our real world test environment.

**Conclusion 2:** *PLC nodes connected to same phase but different breakers operate over lower throughputs as compared to same phase, same breaker case (~20-30% decrease in observed throughput)[2]. This is because signals experience higher attenuations while passing through the breaker circuitry located between the PLC nodes.*

[2]We have excluded the cases of high interference from electric devices.

### D. Impact of Phases

Next, we evaluate the impact of different phases on PLC performance. Overall, the performance degradation when PLC nodes operate on different phases (and different breakers) is similar to the performance degradation when they are placed on the same phase (and different breakers), as shown by the red line on the CDF graph shown in figure 2. The median, minimum and maximum observed jitter was 0.32 ms, 0.15 ms and 1.6 ms respectively. The lowest throughput observed in our experiments is 20 Mbps. This low throughput was caused due to multiple different appliances connected nearby the PLC nodes of that link.

**Conclusion 3:** *PLC nodes operating on different phases and breakers perform similarly to the case where they operate on the same phase but different breaker.*

### E. Impact of Distribution Lines

We finally evaluate the impact of distribution lines on PLC network performance. The transmitter-receiver pairs are connected under different distribution lines (and different breakers and same or different phases). Interestingly, we observe that the connection between PLC nodes is frequently lost and the average round trip times (RTTs) often exceeds 1 second. The maximum throughput that we observed between a pair of nodes N1-N2 was 3 Mbps and 5 Mbps for both directions, and the jitter was varying between 2.03 ms and 5.7 ms. Figures 4(c) and 4(d) show the tonemaps for the 5 different sub-intervals of the AC line cycle. We observe zero modulation for more than half of the subcarriers in N2-N1 direction. The root cause of this poor performance is that PLC signals do not only travel through breaker circuitry but also through the transformer circuitry at the main switch board (MSB) of the building, thus experiencing significantly higher attenuations.

**Conclusion 4:** *PLC performance significantly drops (~ 18-30 folds throughput decrease) when nodes are located at different distribution lines. Distribution lines can make PLC connectivity often impossible.*

## V. ENTERPRISE PLC CHANNEL PROPERTIES & DYNAMICS

In this section, we briefly discuss some properties and channel dynamics of PLC links in enterprise scenario, such as PLC channel reciprocity, impact of appliances on PLC channels and temporal dynamics of PLC links. These discussions form the basis of section VI and VII where we discuss network design strategies and a novel spectrum sharing scheme for PLCs.

### A. PLC Channel Reciprocity

The reciprocity of a PLC link depends on channel frequency response or transfer function between PLC nodes communicating over that link. Asymmetry is attributed to the different multipath characteristics of the powerline, which can vary depending on the location of PLC nodes compared to branch circuits or other connected electrical devices [20], [21], [26]. We quantify asymmetry of a PLC links $a - b$ as:

$$\mathcal{A}_{a,b} = \frac{\sum_{k=1}^{N_{AC}} [\sum_{j=1}^{N} |T_{a \to b}[j]^{\{k\}} - T_{b \to a}[j]^{\{k\}}|]}{N_{AC}} \quad (3)$$

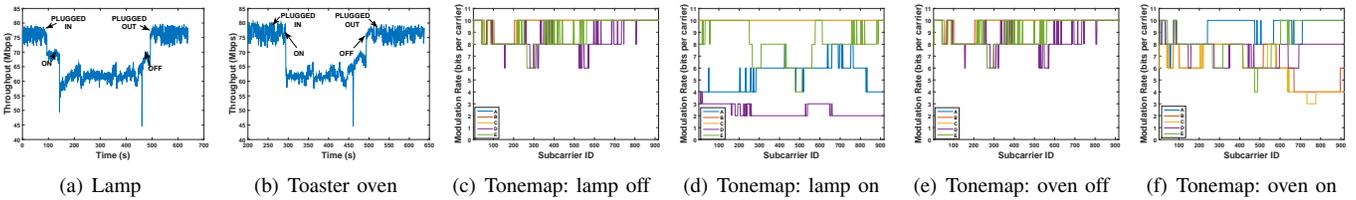

Fig. 5: (a)-(b) Throughput variations for lamp and oven, (c)-(d) Tonemaps with(out) lamp, (e)-(f) Tonemaps with(out) oven

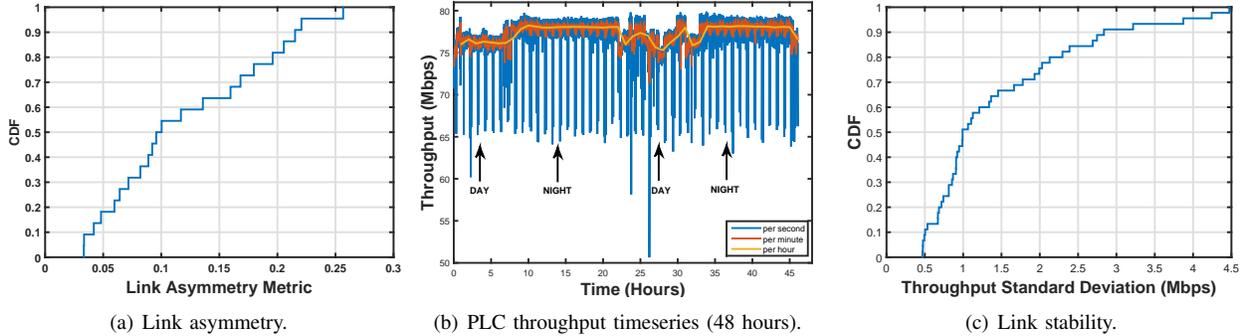

Fig. 6: (a) Link asymmetry, (b) Temporal variation in throughput over 2 days, (c) Link throughput stability CDF (45 links)

where $N$ is the number of subcarriers, $T_j$ is the modulation rate of the $j^{th}$ subcarrier and $N_{AC}$ is the number of sub-intervals of AC line cycle. The above equation estimates asymmetry between two links as the distance between tonemaps of these links, averaged over all AC line cycle sub-intervals. The max and min values for $A_{a,b}$ are 9170 (917 subcarriers $\times$ 10 bits/carrier) and 0, respectively. Please note that our analysis of PLC asymmetric links is different from previous studies such as [5], which only use difference in throughput to quantify PLC asymmetry. Throughput depends not only on PLC link characteristics, but also on other factors such as contention and traffic type. Our results have shown cases where two different links between two same PLC nodes pair have the similar throughput and FEC code rates, but different tonemaps.

In Figure 6(a) we present the distribution of our link asymmetry metric $A_{a,b}$ normalized by the maximum $A_{a,b}$ (which is 9170), from the tonemaps of 25 pair of nodes $a$, $b$. We observe that for more than 50% of the links, the normalized $A_{a,b}$ is greater than 0.1 (917 bits). In terms of measured throughput, the maximum difference observed in asymmetric links is 15 Mbps.

### B. Impact of Appliances on PLC

Electrical devices connected to power lines can impact PLC performance in two ways, i.e., either by introducing adverse multipath attenuations or creating in-band interference. Existing work has shown the impact of such devices on PLC performance by presenting the throughput degradation due to devices connected in between two communicating PLC nodes [7]. We take the analysis of device interference in PLC networks one step further, and show that although different devices cause similar degradation in PLC throughput, their impact on the PLC channel can be very different.

We study the impact of electrical devices on PLC performance by conducting controlled experiments, where we generate back-to-back saturated UDP traffic from one PLC node to another. Figures 5(a) and 5(b) show the PLC throughput variations caused by a lamp and a toaster oven, when they are plugged in, turned on/off and plugged out. We observe that the throughput drops by approximately 15 Mbps for both the lamp and the oven. However, when we analyze the tonemaps for the lamp (figures 5(c)-5(d)) and oven (figures 5(e)-5(f)), we observe that each of these appliances impacts the PLC channel differently. The intensity of interference can also change with the number of connected interfering devices in the power line. As we discussed in Section IV-B, we observed a 84% throughput drop and a significant decrease in per-subcarrier modulation (cf. Figure 3), when multiple devices interfere with PLC communication. Since different PLC links can experience very different channels based on their location with respect to other electrical appliances, dynamic spectrum adaptation can improve throughput performance of PLCs. We design such spectrum adaptation in Section VII.

### C. PLC Temporal Dynamics

Performance of PLCs in enterprise settings can be dynamic either due to interference from already connected appliances, or due to a multitude of electrical devices being turned on/off on a regular basis. In order to study temporal dynamics, we measure performance of a PLC link for a long time periods. Figure 6(b) shows a representative scenario of a PLC link throughput variation, for 2 days (48 hours) period. The throughput variations are averaged over one second, one minute and one hour time windows respectively. We observe that the throughput performance can vary from 52 Mbps to 80 Mbps. The link appears to be highly bursty, which shows that some intense performance dynamics happening at small time scales, which are attributed to interference created by nearby electrical devices. The throughput variations observed at coarser time scales (minutes or hours) are attributed to human activity (e.g. connection/disconnection of new devices, etc.). The analysis of tonemaps (not shown here) also verifies the link variations with time, as we observed that the tonemaps exchanged among PLC nodes during day were different from those during night. However, we observed that throughput between most PLC links remained quite stable. Figure 6(c) shows the CDF plot of standard deviation (averaged over 10 second intervals) of the real time throughput of 45 different links we tested in our building. Throughput for each link was collected over 15 minute time windows. It can be observed that more than 60% of the time, the standard deviation of throughput is below 1.5 Mbps, which shows that throughput performance of most PLC links remains quite consistent over time. This *pseudo-stationary* nature of PLC links can minimize the channel probing overheads, for example, during spectrum sharing (discussed in section VII).

## VI. PLC Network Design for Enterprises

In this section, we discuss (a) network topology design and (b) multi-hop communication for enterprise PLC networks, in the light of measurement study presented in sections IV-V.

### A. PLC Network Topology

The design of a PLC network topology (PLC-Net) is a challenging task both because of the power distribution network characteristics and the interferences from the connected devices in the power line. Our results have shown that PLC performance significantly drops, or connectivity is impossible, when PLC nodes are located at different distribution lines (cf. conclusion 4 - Section IV-E). Hence, PLC-Net should have at least one PLC Internet gateway node for each distribution line. Our experimental floorplan in Figure 1 has 3 distribution lines (indicated hexagon with number 4), and thus requires at least 3 PLC gateways. PLC performance drops when nodes are located at different breakers (cf. conclusion 2 - Section IV-C). On one hand, the deployment of one PLC gateway per breaker can be expensive and often infeasible given that the network (e.g., Ethernet) ports that provide Internet connectivity may not be available close to every breaker. On the other hand, deploying multiple gateways to different breakers per distribution line can allow for PLC devices to dynamically change their gateways based on channel conditions. We leave the examination of this tradeoff as future work.

### B. Case for Multi-hop Routing

HomePlug PLC devices currently do not support multi-hop communication [2], [3]. However, direct link communication in PLC networks can often be impossible or show very low throughput, either due to highly location dependent multipath attenuations and/or interference from appliances (Sections IV-V). In this subsection, we explore if multi-hop routing can improve throughput and connectivity in a large building settings, such as enterprises, through real world experiments.

For our evaluation, we use the optimized link state routing protocol (OLSR) [8], which is a table-driven proactive link-state routing protocol and has been widely used in 802.11 wireless mesh networks. For our testbed experiments, we first port the open-source OLSR and ETT implementations [27], [28] in our OpenWrt boards. Then, we place 9 PLC nodes in various topologies in our floorplan (Figure 1) and then evaluate routing performance of the PLC-Net. Our results show that routing can significantly improve PLC-Net performance in scenarios where certain PLC links perform very poorly. We identified such a scenario during the communication between PLC nodes N9 and N6, which were located at different breakers but in the same distribution line (cf. Figure 1). Figure 7, shows the UDP throughput performance between PLC node N9 and N6 for one minute window, while OLSR is enabled and disabled. When OLSR is turned on, UDP throughput between N9-N6 and N6-N9 is 5.6 and 4.5 times higher, respectively, as compared to the case when OLSR is off. We observe that such communication is affected by electrical devices (lamps, phone chargers, monitors) between N9 and N6, which interfere with the PLC network. When OLSR is enabled, N9 and N6 communicate through node N7 or N8, avoiding such interferences. The throughput temporal variations shown in Figure 7 are attributed to the interference dynamics, which make OLSR to change routes periodically. We make the same observations for TCP traffic (Table I). When OLSR is on, TCP throughput is up to 3.6 times higher compared to the case when OLSR is off.

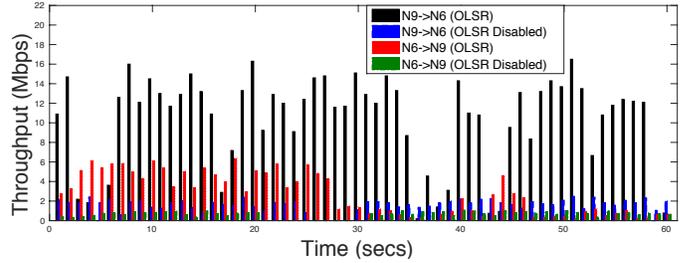

Fig. 7: Throughput with OLSR on/off for 60 secs.

| | | Throughput (Mbps) | Jitter (ms) |
|---|---|---|---|
| UDP | N9→N6 (olsr on) | 9.5 | 5.9 |
| | N9→N6 (olsr off) | 1.7 | 17.8 |
| | N6→N9 (olsr on) | 2.7 | 11.6 |
| | N6→N9 (olsr off) | 0.6 | 18.7 |
| TCP | N9→N6 (olsr on) | 4.2 | - |
| | N9→N6 (olsr off) | 1.4 | - |
| | N6→N9 (olsr on) | 1.8 | - |
| | N6→N9 (olsr off) | 0.5 | - |

TABLE I: UDP and single-flow TCP throughput and jitter with OLSR on/off (jitter is reported by iperf only for UDP).

**Conclusion:** *Mesh routing can significantly boost PLC-Net performance in scenarios where direct PLC links perform very poorly and multi-hop communication is required.*

## VII. Spectrum Sharing in PLCs

HPAV devices contend for the whole spectrum during communication. However, we next show that a fine-grained spectrum management can significantly boost the performance of larger scale PLC networks, such as enterprise PLCs. Our analysis of the tonemaps obtained for several different links in our testbed shows that different PLC links in the same neighborhood can experience significantly different channels, which happens mainly due to highly location dependent multipath characteristics. For example, Figure 8 shows snapshots of the tonemaps of 6 different links in a network of 4 PLC nodes deployed in our test environment. If we consider the last 200 subcarriers (717-917) for all the links of node N1, we observe the modulation is at least 6 bits per carrier (cf. Figures 8(a), 8(b), 8(c)). On the other hand, the last 200 subcarriers for all the links of node N2, show lower modulation, which can be as low as 2 bits per carrier (cf. Figures 8(d), 8(e), 8(f)). A spectrum sharing (SS) strategy could allow both N1 and N2 to transmit at the same time to their neighbors (e.g. N1-N3 and N2-N4) using only their high-performance subcarriers. Similar observations hold for other links (tonemaps not shown here), where certain subcarriers can not carry data (0 modulation) and others can allow high modulations. Next, we present the design of a spectrum sharing solution for PLCs.

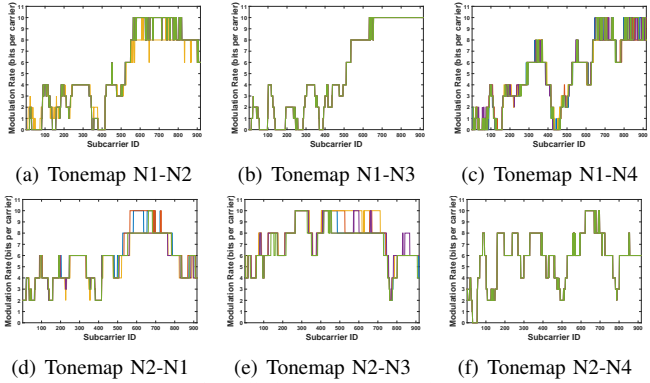

(a) Tonemap N1-N2   (b) Tonemap N1-N3   (c) Tonemap N1-N4
(d) Tonemap N2-N1   (e) Tonemap N2-N3   (f) Tonemap N2-N4

Fig. 8: Tonemaps of 12 links among 4 neighboring PLC nodes.

*A. Spectrum Sharing Strategy*

In this section, we present a spectrum sharing (SS) design, which seeks to improve network throughput. Although, we do not directly seek to provide better fairness, our results show that SS enabled HPAV/AV2 MAC can be more fair in terms of throughput and spectrum allocation. Our proposed design shares the spectrum at OFDM subcarrier level, and can be implemented on existing PLC devices. We design SS for CSMA/CA channel access mode. However, our design can be extended for TDMA based channel access in HPAV. For simplicity, in our discussion we assume that there is one Central Coordinator node (CCo) in the network, and all nodes are in the same collision domain. The CCo determines how the spectrum will be shared among the PLC nodes of its network. CCo separately decides SS strategy for each AC line cycle slot. Moreover, we assume that a snapshot of tonemaps for all PLC links is available at CCo. We further discuss the changes required in HPAV MAC to support SS in Section VII-B.

First, we call the links which occupy the PLC channel through regular HPAV CSMA/CA or TDMA protocol as *primary* (P-Link or $p_{i \to j}$), and the links with which a primary link shares spectrum with, as *secondary* (S-Link or $s_{m \to n}$). For the simplicity of our design, we assume that all links are saturated (i.e., each node always has traffic to send) and whenever a P-Link is established, only one S-Link can operate; i.e. the S-Link which gives maximum possible gain by sharing spectrum with the P-Link. Let $[T_{i \to j}]^p_{1 \times N}$ and $[T_{m \to n}]^s_{1 \times N}$ be the vector of tonemaps of a pair of primary and S-Links $i \to j$ and $m \to n$, respectively. The difference between these two vectors, denoted by $[D_{i \to j, m \to n}]_{1 \times N}$, is:

$$D_{i \to j, m \to n} = T_{m \to n} - T_{i \to j} \quad (4)$$

Based on the difference calculated from eq. (4), the CCo can determine which subcarriers perform poorly as compared to the subcarriers of the S-Link, and allocate those subcarriers to the S-Link which will lead to maximum throughput gain.

Let $\widetilde{D}_{i \to j, m \to n}$ denote all the elements in $D_{i \to j, m \to n}$ which are above a certain threshold $\beta$; i.e. $\widetilde{D}_{i \to j, m \to n} = \{\forall l \in D_{i \to j, m \to n} | l >= \beta\}$. Let $I_{m \to n}$ be the set of indices corresponding to the values of those subcarriers. The gain $G_{m \to n}$ obtained by allowing an S-Link $q_{m \to n}$ to operate along with a P-Link is:

$$G_{m \to n} = \sum_{I_{m \to n}} [T_{m \to n}] - \sum_{I_{m \to n}} [T_{i \to j}] \quad (5)$$

CCo selects the S-Link which maximizes the gain $G_{m \to n}$. It then communicates the indices $I_{m \to n}$ to the transmitting node of the best S-Link and the P-Link. The P-Link/S-Link will disable/enable the subcarriers corresponding to these indices during SS. The throughput gains of SS can be approximated using eq. (2) as $Thr_p^{N-I} + Thr_s^I$, where $Thr_s^I$ is the throughput when $I$ subcarriers are assigned to S-Link $s$.

In practice, the best secondary node might not have any traffic to send. To avoid underutilization of spectrum, CCo can select and share SS information with top $M$ candidate secondary nodes, which can all contend for the spectrum shared by the P-Link. Although, we did not observe such cases in our deployments, but in practice, a PLC-Net can also contain some extremely bad links (modulation of all subcarriers is very low). In order to prevent such bad links from starving, each node can limit its SS and choice of $\beta$, such that it only shares a certain percentage of its subcarriers during SS. Intuitively, a bad link should share smaller percentage of its spectrum, and vice versa.

Our proposed SS works on top of regular HPAV CSMA/CA procedure [12], [17]. SS is performed only when a P-Link is established and is already operating. We next elaborate on how our proposed SS can work in concert with HPAV.

*B. Proposed Spectrum Sharing Mechanism*

In the following steps, we present a mechanism through which the current HPAV MAC and PHY layers can support the SS strategy proposed above.

**1: Distribution of SS decisions by CCo.** CCo of a PLC-Net will periodically log tonemaps of all possible links to formulate SS decisions according to the SS strategy VII-A. The tonemaps logged by different nodes in the network, during periodic full spectrum transmissions, will be used by CCo for re-evaluation of PLC links. CCo can use Management Messages (MMEs) for this purpose [2], [3]. CCo will choose the probing interval $\tau_m$ such that the MMEs incur minimal overhead and interference to data transmissions. Particularly, in cases when PLC channels are very dynamic, probing frequency $\frac{1}{\tau_m}$ will always be less than some threshold (which can be chosen according to the temporal dynamics of the links in its network V-C). Note that the CCo does not always need to probe for the tone maps of all the links. This is because CCo monitors the traffic over its network, and therefore, it can have tonemaps information of some links in advance. Note that SS will not be performed during the exchange of MMEs.

After processing the tonemaps, CCo will distribute its SS decisions to the nodes in its network, such that each node receives SS allocation information with every other node with which it can form a possible link. The whole process of

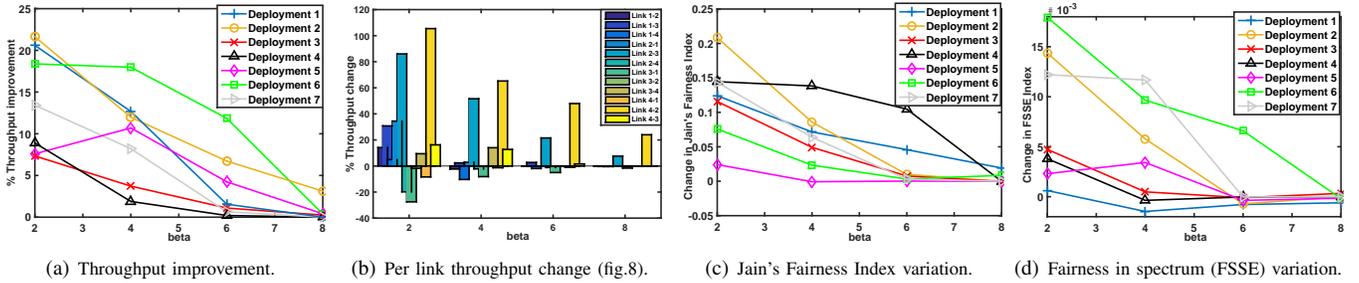

(a) Throughput improvement. (b) Per link throughput change (fig.8). (c) Jain's Fairness Index variation. (d) Fairness in spectrum (FSSE) variation.

Fig. 9: Spectrum Sharing throughput and fairness performance for seven 4-node PLC deployments in test environment

probing and distribution of decisions will require CCo to send $2 \times (N-1)^2$ messages of varying lengths every $\tau_m$ seconds, where $N$ = number of nodes in the network including the CCo. At the end of this step, each node in the network will know which part of the spectrum to use during SS if they establish a P-Link towards any other node in the network. Moreover, each node will know its rank among the top $M$ candidate secondary nodes corresponding to each possible P-Link in the PLC-Net.

**2: Medium access during SS.** Current HPAV MAC is centrally controlled through beacon signals from CCo. CCo broadcasts robust and reliable beacons, to establish Beacon Periods (BPs) consisting of TDMA and CSMA slots. Here, we focus on the CSMA slots only. Note that in case of TDMA, the CCo can simply schedule the P-Link and one of the top $M$-best performing S-Links to operate in the same slot.

As we mentioned before, even with SS enabled, all nodes in a PLC-Net will always be in contention for the whole spectrum, globally throughout the network. However, the medium access and the consequent link interactions during SS will differ from the regular HPAV CSMA/CA as following:

**(i)** After a P-Link is established, the remaining nodes will go into their backoff stages, following global CSMA/CA procedure. However, for SS, the primary node will also enable the *Multicast Flag* (MCF) in the *Start-of-Frame Control* (SOF) field of its *MAC Protocol Data Unit* (MPDU) [25], so that all remaining nodes in the network can extract the source and destination IDs of the P-Link from this SOF delimiter field. From then onwards, nodes of the established P-Link will negotiate with each other, to use only the unshared subcarriers for transmission, while disabling the shared subcarriers.

**(ii)** After knowing the link information from SOF delimiter in P-Link's broadcast MPDU frame, nodes of the corresponding top $M$-best S-Links will disable the subcarriers being used by their P-Link. These candidate S-Links will then wait for short intervals, proportional to the ascending order of their ranks, to let the better links to operate first. After establishment of the first S-Link, all $M$ S-Links will go into CSMA/CA contention over the shared spectrum. This contention will only be applicable during SS and will be managed independently from the global contention for the whole spectrum.

**(iii)** If the backoff counter (corresponding to the global CSMA/CA procedure) of any node expires, it will try transmitting across the whole spectrum, to initiate its own P-Link. If that node belongs to one of the active $M$-best S-Links, it will stop its previous transmission and re-enable all the subcarriers before trying transmission across the whole spectrum.

**(iv)** Nodes belonging to any active P-Link will come out of their SS state (i.e. re-enable all subcarriers and enter into global contention for the whole spectrum) if they hear an attempted transmission in the unshared part of the spectrum, or, if the transmission between them is finished. As soon as transmission over P-Link ends, the S-Links corresponding to that P-Link will stop sensing transmission in the unshared part of the spectrum, and immediately come out of their SS state.

**(v)** The aforementioned process (i)-(iv) repeats for every new P-Link in the PLC-Net.

**3. Periodic re-evaluation of full spectrum:** All nodes in the network will periodically decide to disable SS and transmit across full spectrum. No S-Link will operate in this case. The frequency of this periodic behavior can be either chosen by each node independently or by CCo of the network, based on local or global temporal dynamics V-C). Such periodic use of the whole spectrum will allow each node to automatically update its full spectrum tonemaps towards other nodes in the network, during regular data transmissions. The network CCo will then re-evaluate the SS decisions in its network by accessing these tonemaps as described before.

Note that SS strategy also applies to the preambles of HPAV frames exchanged between the nodes (i.e. the preambles exchanged between nodes of the established links are transmitted only over the allocated spectrum). However, SS will not be performed during the establishment of a P-Link.

*C. Evaluation*

We perform trace driven simulations using the traces collected from seven different 4-node PLC deployments (Figure 8 represents Deployment#1). Our simulation does not take into account frame aggregation procedures, bit loading of ethernet frames inside PLC frames, management messages and channel errors, since these parameters are proprietary vendor-specific implementation information. In our simulations, we choose collision duration $T_c = 2920.64\mu s$, duration of successful transmission $T_s = 2542.64\mu s$ and frame length $F_l = 2050$ [12], [29]. The contention window (CW) and deferral counter (DC) values used for each HPAV CSMA/CA back off stage are [8, 16, 32, 64] and [0, 1, 3, 15], respectively.

**Throughput gains.** We calculate the normalized throughput $Thr$ for each link $m \rightarrow n$ in our simulation as follows:

$$Thr = 100 \cdot \frac{[\sum_{i=1}^{[\#SuccessTransmissions]} S_{F_i}] \cdot [Frame\ length]}{Total\ simulation\ time}$$

$S_{F_i}$ represents the fraction of spectrum utilized at $i$-th transmission. $S_{F_i} = \sum_{j=1}^{N}[T_{m \to n}]/9170$, such that $\max(S_{F_i}) = 1$ and $\min(S_{F_i}) = 0$, where $T_{m \to n}$ are tonemaps of the link $m \to n$.

Figure 9(a) shows the effect of varying SS parameter $\beta$ on the overall network throughput (percentage increase) for seven different deployments. We observe an improvement in overall throughput for all seven deployments. Moreover, we observe that the throughput gains increase when $\beta$ decreases, because smaller $\beta$ allows for more low-performance subcarriers to be assigned to S-Links. Figure 9(b) shows the per-link percentage throughput improvement for the example mentioned in Figure 8 (Deployment#1). We observe that the normalized throughput of 7 out of 12 links is increased for $\beta = 2$. However, we also observe a decrease in throughputs of the remaining links, since those links share more spectrum as compared to others. Links such as 2-3, 4-2 and 4-3 experience higher gains as compared to others. This is because whenever the spectrum is shared with these links, it happens to be in the part of spectrum of these links, where modulation index is high. In this example, the highest aggregated percentage *Thr* gains exceeded 20% (blue line in Figure 9(a)), while per-link throughput gains were as high as 104% (Link 4-2, Figure 9(b)).

**Improvement in fairness.** We evaluate the fairness of our spectrum sharing strategy, by calculating *Jain's fairness index* (JFI) [30] and *Fairly Shared Spectrum Efficiency* (FSSE) [31] for the above mentioned deployments. An allocation strategy is maximally fair if all nodes in a PLC-Net allocate the same throughput, in which case JFI = 1. On the other hand, FSSE of a PLC-Net gives the spectrum efficiency of the PLC node with minimum throughput in the network. In case of maximum spectrum fairness, FSSE is equal to the spectrum efficiency of the whole network [31].

Figures 9(c) and 9(d) show the difference in fairness between SS-enabled and SS-disabled for JFI and FSSE, for the seven PLC-Net deployments we tested. Interestingly, we observe that SS leads to an increase of both JFI and FSSE indexes. Furthermore, fairness decreases as we increase $\beta$ (and vice versa). These results show that spectrum sharing in PLCs can not only increase the overall throughput, but also the fairness in throughput and spectrum allocation for PLC-Nets.

## VIII. CONCLUSIONS

PLC technology has the potential to improve connectivity and allow for new applications in enterprise settings at low cost, without any need for dedicated network cabling. In this paper, we first conduct a measurement study using HomePlug AV testbeds, to explore PLC performance in large buildings such as enterprises. Our results uncover the impact of the power distribution network components (such as breakers, distribution lines, AC phases) and electrical interference on PLC performance, in spatial, temporal and spectral dimensions. The key insight gained is that careful topology planning with multiple PLC gateway nodes per distribution line and routing are important requirements for robust connectivity in PLC networks. Furthermore, fine-grained spectrum sharing (which is not supported by existing PLCs) can significantly boost performance of PLC-Nets, both in terms of increased aggregated throughput as well as fairness. We expect that our study will stimulate more community effort on bringing PLCs to large buildings (e.g. enterprises, university campuses, etc.).

## REFERENCES

[1] Homeplug technology gains momentum. In *http://www.businesswire.com*, 2013.
[2] Homeplug av whitepaper. *http://www.homeplug.org/tech-resources/resources/*, 2007.
[3] Homeplug av2 whitepaper. *http://www.homeplug.org/tech-resources/resources/*, 2011.
[4] Pat Pannuto and Prabal Dutta. Exploring powerline networking for the smart building. In *Ann Arbor 1001: 48109*, 2011.
[5] Christina Vlachou et al. Electri-fi your data: Measuring and combining power-line communications with wifi. In *Proc. of ACM Internet Measurement Conf.*, number EPFL-CONF-211905, 2015.
[6] Ahmed Osama Fathy Atya et al. Bolt: Realizing high throughput power line communication networks. In *Proc. of ACM CoNEXT*, 2015.
[7] Rohan Murty et al. Characterizing the end-to-end performance of indoor powerline networks. *Harvard University Microsoft Research*, 2008.
[8] Clausen et al. Optimized link state routing protocol (olsr). In *RFC 3626*, 2003.
[9] Richard Nizigiyimana et al. Characterization and modeling breakers effect on power line communications. In *IEEE Int. Symp. on Power Line Communications and Its Applications*, 2014.
[10] Shinji Tsuzuki, Utsunomiya Hiroaki, and Yamada Yoshio. One wire plc system for inter-floor connectivity. In *IEEE Int. Symp. on Power Line Communications and its Applications*, 2014.
[11] Ogawa et al. A study on application of power line communication to lan in building. In *IEEJ Transactions on Electrical and Electronic Engineering*, 2008.
[12] Christina Vlachou et al. On the mac for power-line communications: Modeling assumptions and performance tradeoffs. In *IEEE Int. Conf. on Network Protocols (ICNP)*. IEEE, 2014.
[13] Christina Vlachou et al. Fairness of mac protocols: Ieee 1901 vs. 802.11. In *IEEE Int. Symp. on Power Line Communications and Its Applications*. IEEE, 2013.
[14] Mauro Biagi et al. Location assisted routing techniques for power line communication in smart grids. In *IEEE Int. Conf. on Smart Grid Communications*, 2010.
[15] Mauro Biagi et al. Neighborhood-knowledge based geo-routing in plc. In *IEEE Int. Symp. on Power Line Communications and Its Applications*, 2012.
[16] Kaveh Razazian et al. Enhanced 6lowpan ad hoc routing for g3-plc. In *IEEE Int. Symp. on Power Line Communications and Its Applications*, 2013.
[17] Ieee standard for broadband over power line networks: Medium access control and physical layer specifications. In *IEEE Std. 1901*, 2010.
[18] Matthias Gtz, Rapp Manuel, and Dostert Klaus. Power line channel characteristics and their effect on communication system design. *IEEE Communications Magazine*, 2004.
[19] Holger Philipps. Modelling of powerline communication channels. In *Proc. of Int. Symp. on Power-Line Communications and its Applications*, 1999.
[20] Manfred Zimmermann and Dostert Klaus. Analysis and modeling of impulsive noise in broad-band powerline communications. *IEEE Transactions on Electromagnetic Compatibility*, (1), 2002.
[21] Manfred Zimmermann and Dostert Klaus. A multipath model for the powerline channel. *IEEE Transactions on Communications*, (4), 2002.
[22] D. Chariag et al. Channel modeling and periodic impulsive noise analysis in indoor power line. In *IEEE Int. Symp. on Power Line Communications and Its Applications*. IEEE, 2011.
[23] Luca Di Bert et al. On noise modeling for power line communications. In *IEEE Int. Symp. on Power Line Communications and Its Applications*. IEEE, 2011.
[24] Hasan Basri elebi. *Noise and multipath characteristics of power line communication channels*. PhD thesis, University of South Florida, 2010.
[25] Latchman et al. *Homeplug AV and IEEE 1901: A Handbook for PLC Designers and Users*. John Wiley & Sons, 2013.
[26] Manfred Zimmermann and Dostert Klaus. An analysis of the broadband noise scenario in powerline networks. In *Int. Symp. on Powerline Communications and its Applications*, 2000.
[27] Olsr routing protocol. In *http://www.olsr.org*.

[28] Olsr with link cost extensions. In *http://sourceforge.net/projects/olsr-lc/*.
[29] Christina Vlachou et al. Analyzing and boosting the performance of power-line communication networks. In *Proc. of ACM Int. on Conf. on emerging Networking Experiments and Technologies*. ACM, 2014.
[30] Raj Jain and *et al.*. *A quantitative measure of fairness and discrimination for resource allocation in shared computer system*. Eastern Research Laboratory, MA, 1984.
[31] Magnus Eriksson. Dynamic single frequency networks. *Selected Areas in Communications, IEEE Journal on*, 2001.